\documentclass[aps,prl,superscriptaddress,showpacs,floatfix,twocolumn]{revtex4}
\usepackage{graphicx}	

\def\dAu  {$d$+Au}
\def\AuAu {Au+Au}
\def\pp   {$p+p$}

\def\AA   {$A+A$}
\def\AB   {$A+B$}

\def\sqrtsNN {\mbox{$\sqrt{s_{NN}}$}}

\def\Ncoll   {\mbox{$N_{\rm coll}$}}

\def\RdAu    {\mbox{$R_{d\rm Au}$}}

\def\jpsi    {\mbox{$J/\psi$}}
\def\pT      {\mbox{$p_{T}$}}
\def\beq     {\begin{equation}}
\def\eeq     {\end{equation}}

\begin{document}

\title{Transverse momentum dependence of \jpsi\ shadowing effects}
\author{
E. G. Ferreiro} 
\affiliation{Departamento de F\'{\i}sica de Part\'{\i}culas, 
Universidade de Santiago de Compostela, Spain} 
\affiliation{Laboratoire Leprince Ringuet, \'Ecole Polytechnique,
91128 Palaiseau, France} 
\author{F. Fleuret} 
\affiliation{Laboratoire Leprince Ringuet, \'Ecole Polytechnique,
91128 Palaiseau, France} 
\author{A. Rakotozafindrabe} 
\affiliation{IRFU/SPhN, CEA Saclay, 91191 Gif-sur-Yvette Cedex, France}
\affiliation{Laboratoire Leprince Ringuet, \'Ecole Polytechnique,
91128 Palaiseau, France} 

\begin{abstract}
We present a new approach to estimate the effect of the gluon shadowing in nucleus+nucleus collisions and its consequences 
on the \jpsi\ production yield. Using kinematical information available from the measured \jpsi\ production in proton+proton collisions 
at \sqrtsNN\ = 200~GeV, we build a 
Glauber 
Monte Carlo code which takes into account shadowing in two alternative ways: multiple scattering 
corrections or Q$^2$ evolution of parton densities. We exploit the dependence of these different parameterizations to the \jpsi\ 
transverse momentum and we give the first predictions on the resulting $p_T$ dependence of the nuclear modification factor in 
deuteron+gold collisions at the same energy.
\end{abstract}

\maketitle

The transition from hadronic matter to a deconfined state of quarks and gluons, the so-called Quark-Gluon Plasma (QGP), 
is the subject of intense experimental and theoretical works in heavy ion physics. 
Recent results on \jpsi\ production measured by the PHENIX experiment at the BNL Relativistic Heavy Ion Collider show 
a significant suppression of the \jpsi\ yield in \AuAu\ collisions at \sqrtsNN=200~GeV \cite{Adare:2006ns}. 
A possible origin of this suppression is the screening of the charmonium pair in the hot and dense nuclear medium, 
as expected in case of a QGP formation. 
Nevertheless, PHENIX data on $d$+Au collisions \cite{Adare:2007gn} have also revealed that cold nuclear matter effects play an essential 
role at these energies. 
In fact, in order to determine the importance of hot and dense matter effects on \jpsi\ production in \AA\ collisions, 
it is fundamental to have a good understanding and a proper baseline for cold nuclear matter effects including initial-state 
shadowing corrections.

Motivated by this goal, we have developed a Monte Carlo code, based on the Glauber model, where the cold nuclear matter 
effects - {\it shadowing} - have been introduced. The inputs of our Monte Carlo are the \jpsi\ rapidity and 
transverse momentum distributions extracted from \sqrtsNN=200~GeV \pp\ data measured by the PHENIX experiment \cite{Adare:2006kf}.
We investigate two different possibilities for the shadowing corrections: on one side we consider a multiple scattering 
approach, where the coherence effects lead to a modification of the nuclear ratio; on the other side we apply a model based on
$Q^2$-evolution of the nuclear ratios of parton distribution functions (PDF).

We present our results for both models on the nuclear modification factor \RdAu\ as a function of rapidity $y$,
number of collisions \Ncoll\ and transverse momentum \pT.
For the first time, we present the \pT\ dependence of the shadowing effects on the \jpsi. We found significant differences
according to the type of shadowing considered, hence giving a new opportunity to test shadowing schemes. 
Note that, in order to show up shadowing effects, we
have not included the contribution from the nuclear absorption, i.e. the \jpsi\ break-up by nucleons in the final state.  

Shadowing refers to the mechanism that makes the nuclear structure functions in nuclei different 
from the superposition of those of their constituents nucleons. 
Several explanations have been proposed. 
Here, we will consider two different approaches :
\begin{itemize}
\item On one hand, in the rest frame of the nucleus, nuclear shadowing can be understood as a consequence of multiple 
scattering \cite{Capella:1999kv,Capella:2005cn}, the incoming virtual photon splitting into a colorless 
$q\bar{q}$ dipole which interacts with the nucleus with typical hadronic cross sections resulting in 
absorption. 
\item On the other hand we will consider the evolution models \cite{Eskola:1998df}. These approaches provide the PDF at 
a given initial value of $Q^2$ and study its evolution through the Dokshitzer-Gribov-Lipatov-Altarelli-Parisi (DGLAP) 
equations. 
\end{itemize}
Following the first approach - models based on multiple scattering - \cite{Capella:2005cn}, one can calculate the 
reduction factor due to the shadowing corrections at fixed impact parameter $b$ for a nucleus as
\beq
\label{eq1}
R^A_{sh}(b,y,p_T)=\frac{1}{1+A\ T_A(b)\ F(y,p_T)}\ .
\eeq
The function $F(y,p_T)$ is given by the integral of the ratio of the triple Pomeron cross-section over
the single Pomeron one - equivalent to the reduction due to the interaction among the gluons - 
\begin{eqnarray}
\label{eq2}
F(y,p_T) & = & \left. 4 \pi \int_{y_{min}}^{y_{max}} dy \frac{1}{\sigma^{P}} \frac{d^2 \sigma^{PPP}}{dy dt} \right |_{t=0} \nonumber \\
         & = & C \left [\exp \left (\Delta y_{max}\right ) - \exp \left ( \Delta y_{min}\right )\right ],
\end{eqnarray}
where $y_{min}=\ln{\left(\frac{R_A m_N}{\sqrt{3}}\right)}$ and $y_{max}=\frac{1}{2}\ln{\left(\frac{s}{m_T^2}\right)} \mp y$.
$y$ is the center of mass rapidity of the produced particle, $y>0$ for the projectile hemisphere and $y<0$ for 
the target one. $m_T$ is the transverse mass, $m_T=\sqrt{m^2+p_T^2}$. 
The parameters $C=0.31$ fm$^2$ and $\Delta=0.13$ 
are fixed through data on DIS scattering \cite{Capella:1999kv}. In the following we will refer to this model as CF shadowing.

The second type of models study the $Q^2$-evolution of nuclear ratios of PDF,
\beq
\label{eq4}
R^A_i (x,Q^2) = \frac{f^A_i (x,Q^2)}{ A f^{nucleon}_i (x,Q^2)}\ , \ \
f_i = q, \bar{q}, g,
\eeq
through the DGLAP evolution equations. 
Nuclear ratios are parameterized at some initial scale $Q^2_0 \sim 1 \div 2$ GeV$^2$ which is assumed large enough for perturbative 
DGLAP evolution to be applied reliably. 
These initial parameterizations for every PDF have to cover the full
$x$ range: $0 < x < 1$. Then these initial conditions are evolved through the DGLAP equations towards
larger values of $Q^2$ and compared with experimental data. 
From this comparison the initial parameterizations are adjusted.
The centrality dependence of shadowing is not addressed in these models as the existing experimental data do not allow its precise  
determination, although some approaches provide an ansatz for such a dependence.
It can be parameterized \cite{Vogt:2004dh} assuming that the inhomogeneous shadowing is proportional to the local 
density, $\rho_A(b,z)$, 
\beq
\label{eq5}
R^A_i (b,x,Q^2)=1+[R^A_i (x,Q^2)-1]N_{\rho}
\frac{\int dz \rho_A(b,z)}{\int dz \rho_A(0,z)}\ ,
\eeq
where $b$ and $z$ are the transverse and longitudinal location in position space, $\rho_A(b,z)$ corresponds
to the Woods-Saxon distribution for the nucleon density in the nucleus, related to the nuclear profile function $T_A(b)$ by 
$\int dz \rho_A(b,z)= A\ T_A(b)$, $\rho_0$ is the central density, given by the normalization $\int d^2b \int dz \rho_A(b,z) = A$
and $R^A_i (x,Q^2)$ is the shadowing function from EKS98 as defined by eq. (\ref{eq4}).
The integral over $z$ includes the material traversed by the incident nucleon, so we are considering that the 
incident parton interacts coherently with all the target partons along its path length. $N_{\rho}$ is the normalization 
factor, and it is chosen so that $(1/A) \int d^2b \int dz \rho_A(b,z) R^A_i (b,x,Q^2) = R^A_i (x,Q^2)$.
In the following we will refer to this model as EKS shadowing.

Within our framework, we study the modification of the \jpsi\ production in nuclear matter. It is often presented 
using the nuclear modification factor:
\begin{equation}
R_{AB} = \frac{dN_{AB}^{J/\psi}}{\langle N_{coll}\rangle dN_{pp}^{J/\psi}}
\end{equation}
where $dN_{AB}^{J/\psi}$ and $dN_{pp}^{J/\psi}$ are respectively the \jpsi\ yield observed in \AB\ collisions
and in \pp\ collisions, and $\langle N_{coll}\rangle$ is the average number of nucleon-nucleon collisions occurring
in a \AB\ collision. For a hard process, such as \jpsi\ production, in absence of nuclear effects, the $R_{AB}$
ratio should equal unity. 

In the present study, this quantity is obtained with a Glauber Monte Carlo. Within this code we describe the collision of 
two nuclei, determine the number of nucleons in the path of each incoming nucleon and calculate the number of 
nucleon-nucleon collisions. Nuclear density profiles have been defined using a Wood-Saxon parametrization for 
any nucleus $A>2$ and the Hulthen wavefunction for the deuteron
\cite{Hodgson:1971}. The nucleon-nucleon inelastic cross-section at
\sqrtsNN~=~200~GeV has been taken to 42 mb and the average nucleon density to 0.17 nucleons/fm$^3$.

In absence of nuclear effects, the \jpsi\ yield in a nucleus-nucleus collision corresponds to a simple superposition of 
the yield obtained in \pp\ collisions multiplied by the average number of nucleon-nucleon collisions. In \pp\
collisions, the \jpsi\ production has been measured at \sqrtsNN~=~200~GeV by the PHENIX experiment, providing the 
\jpsi\ differential cross section versus both the transverse momentum \pT\ and the rapidity $y$ \cite{Adare:2006kf}. 

Based on the parameterizations used to fit these two distributions we can generate a sample of \jpsi: 
we randomly pick $y$, \pT\ and $\phi$ (the azimuthal angle, described as a flat distribution within [0,2$\pi$]) 
and use these values as the inputs of our Monte Carlo. Then, using the shadowing parametrization, we can 
compute $R_{\rm shadow}^{AB}$, the nuclear modification factor when the \jpsi\ production is affected by shadowing. 

\begin{figure}[thb]
  \includegraphics[width=0.49\linewidth, height=0.4\linewidth]{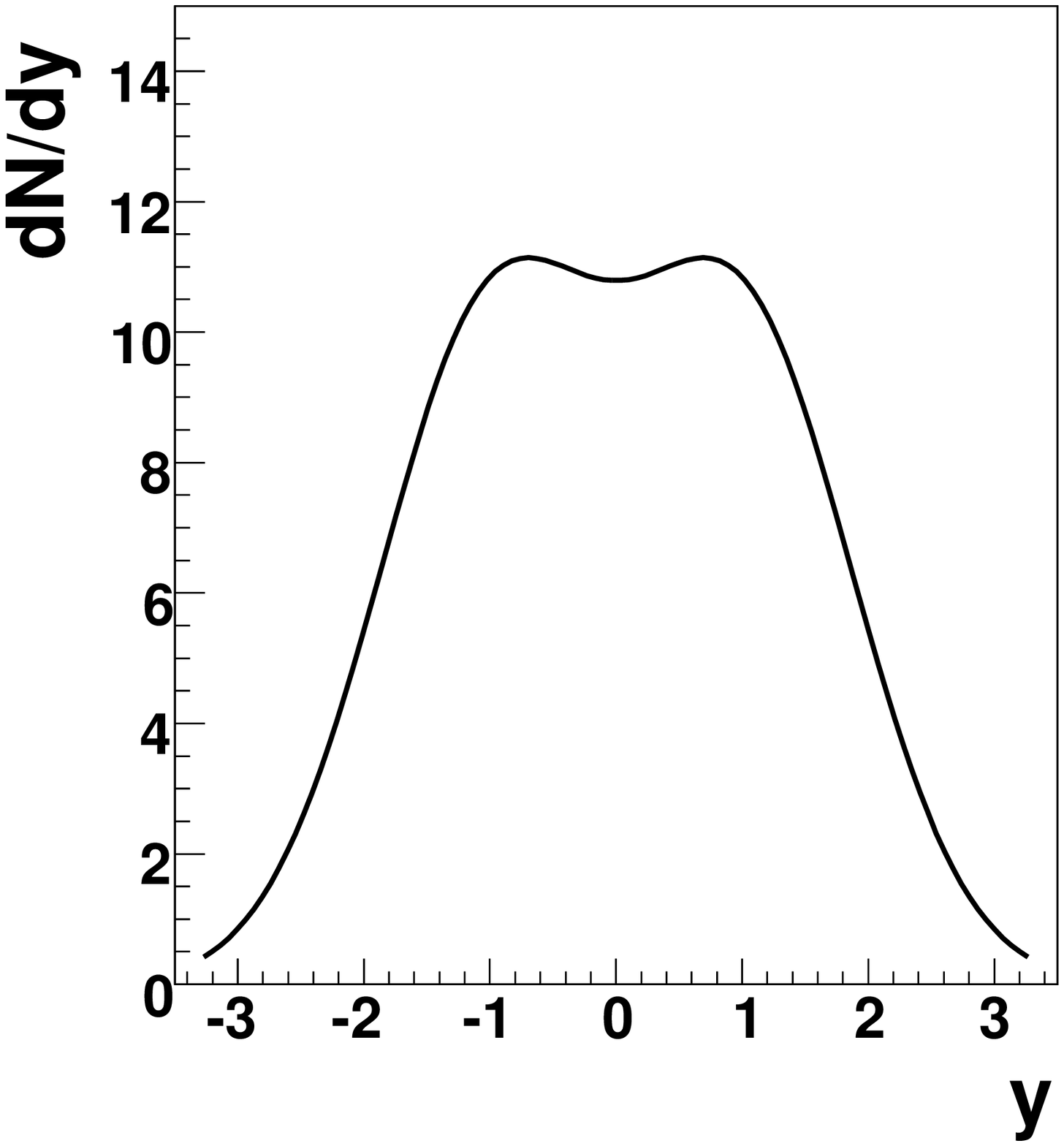}  
  \includegraphics[width=0.49\linewidth, height=0.4\linewidth]{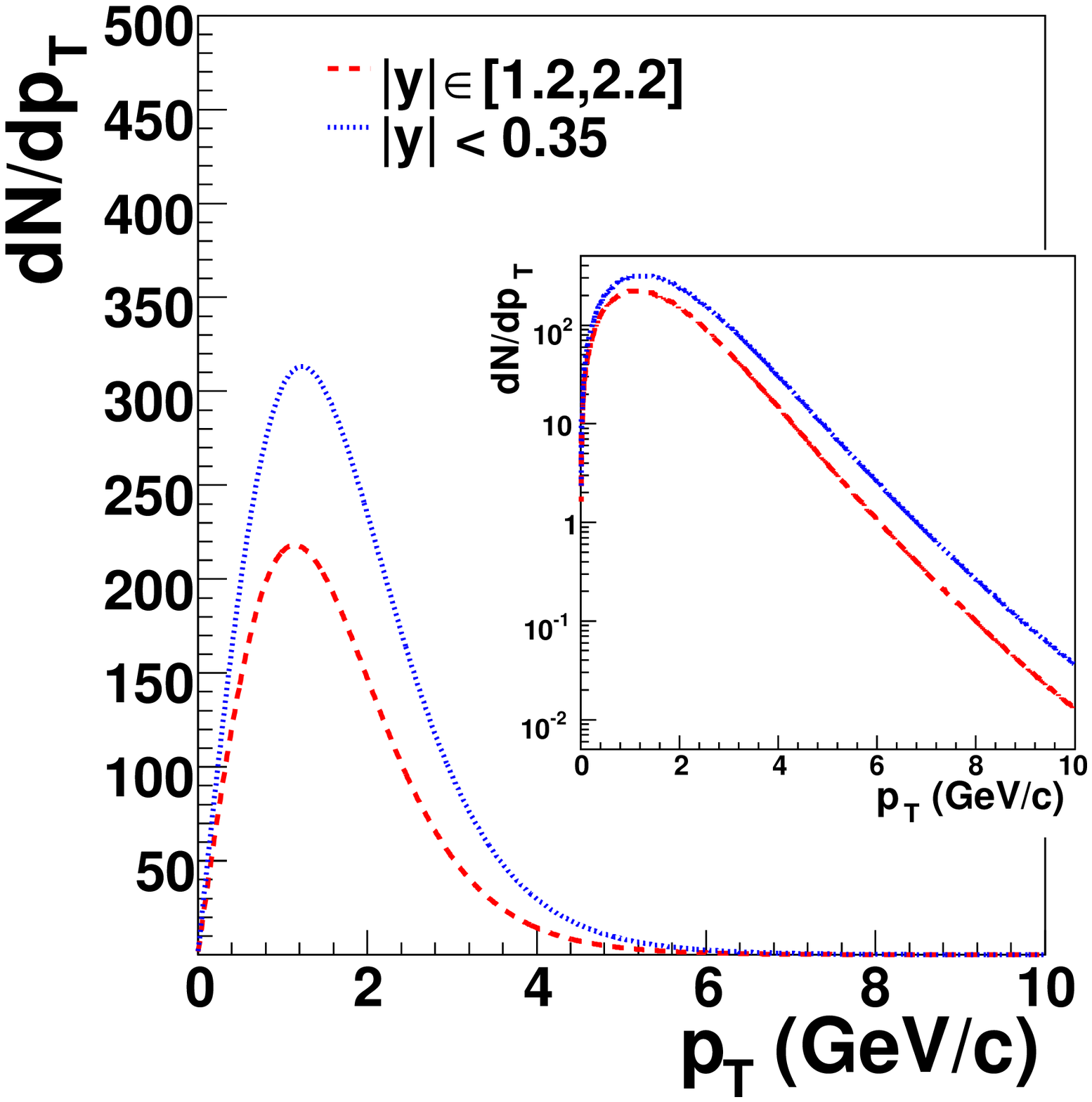} 
  \caption{\label{fig:input variables}
    \jpsi\ rapidity and \pT\ distributions in \pp\ collisions extracted from \cite{Adare:2006kf}. Vertical scales are arbitrary units.}
\end{figure}

Figure \ref{fig:input variables} shows the distributions used in our code. Note that in order to 
restrain ourself to a physical phase space domain, we require for each \jpsi\ the kinematic conditions $0<x_1,x_2<1$ where
\begin{eqnarray}
  x_1 = \frac{m_T}{\sqrtsNN}e^{y} \hspace{0.5cm} {\rm and} \hspace{0.5cm} x_2 = \frac{m_T}{\sqrtsNN}e^{- y}\ .
\label{eq:x12}\end{eqnarray}

As previously mentioned, the shadowing models predict that the modification of the structure functions depends on the 
nuclear environment surrounding the interacting partons. They both 
parameterize this spatial dependence as a function of $A\ T_A(b)$ where $A$ is the nucleus mass
number and $T_A(b)$ is the nuclear profile function at fixed impact parameter $b$, normalized to unity. In fact, $A\ T_A(b)$
corresponds to the number of nucleons seen by a given nucleon at a distance $b$. This quantity depends on
the position of the given nucleon in the nucleus; central nucleons see more surrounding nucleons than
peripheral ones. Within our framework, this number can be computed event by event, thus taking into 
account the spatial nucleon distribution for the current nucleus. Defining as $n_a$ the nucleon which 
suffers shadowing and $d$ the distance between $n_a$ and a nucleon $n_b$ belonging to the same nucleus, 
$n_a$ suffers shadowing from $n_b$ if $\pi d^2 < \sigma_{tr}$, where $\sigma_{tr}$ corresponds to
the nucleon transverse area. Since the average nucleon density in a nucleus is 
$\rho_0=(\frac{4}{3}\pi r^3)^{-1}=0.17$ nucleons/fm$^3$, we take $\sigma_{tr}=\pi r^2=3.94$ fm$^2$. 
We can then compute $N_{tr}$, that is, the number of nucleons which add a shadowing contribution to $n_a$.
Finally, $R_{\rm shadow}^{AB}$ is computed following respectively \cite{Capella:2005cn} and \cite{Vogt:2004dh} for CF and EKS shadowing. 

In the context of the CF model, $R_{\rm shadow}^{AB}$ is computed using the Schwimmer expression given in Eq.~(\ref{eq1}). For \AB\
collisions, we get:
\begin{eqnarray}
  R_{\rm shadow}^{AB}& = & R_{\rm shadow}^A \times R_{\rm shadow}^B  \nonumber \\
            & = & \frac{1}{1+N_{tr}^A F(y,p_T)}  \times\frac{1}{1+N_{tr}^B F(y,p_T)} 
\end{eqnarray}
where $F(y,p_T)$ is the CF shadowing function defined in Eq.~(\ref{eq2}), and $y$ and \pT\ are the rapidity and the transverse momentum 
of the \jpsi\ candidate.

In the EKS calculation \cite{Eskola:1998df}, the authors
provide the ratio $R_i^A$ of the PDF in a proton of a nucleus $A$ to the PDF in the free 
proton. They give a numerical parametrization of $R_i^A(x,Q^2)$ for all parton flavours.
Here, we restrain our study to gluons since, at high energy, \jpsi\ is essentially produced through 
gluon fusion \cite{Lansberg:2006dh}.
The numerical parametrization is valid for: $A>2$ (otherwise return 1) where $A$ is the atomic number,
$10^{-6} < x < 1$ and $Q^2>2.25$ GeV$^2$. Within our framework, we used the table provided in \cite{Eskola:1998df} 
where the inputs are
$A$, $x$ and $Q^2$ and the output is $R_i^A(x,Q^2)$ which gives homogeneous shadowing. In order
to include inhomogeneous shadowing, meaning that shadowing should depend on the interacting 
parton spatial position, we follow the prescription introduced in
\cite{Vogt:2004dh} and define $R_{\rm shadow}^{AB}$
as:
\begin{eqnarray}
  R_{\rm shadow}^{AB} & = & R_{\rm shadow}^A \times R_{\rm shadow}^B  \nonumber \\
                  & = & \left(1+[R_g^A(x_1,Q^2)-1]\times\frac{N_{tr}^A}{\langle N_{tr}^A\rangle}\right
) \nonumber \\
            & \times & \left(1+[R_g^B(x_2,Q^2)-1]\times\frac{N_{tr}^B}{\langle N_{tr}^B\rangle}\right
)
\end{eqnarray}
where $N_{tr}^{A(B)}/\langle N_{tr}^{A(B)}\rangle$ reflects the local nuclear density. Note 
that we implicitly assume here that the incident parton interacts coherently with all 
the target partons in its path.\\

The $x_1$ and $x_2$ values are obtained from $y$ and \pT\ as in Eq.~(\ref{eq:x12}). Doing so is equivalent to consider 
the hard process $g+g\rightarrow c\bar{c}$, where the $c\bar{c}$ nonzero \pT\ is inherited from the initial gluon
transverse momenta. For the square of the momentum transfer $Q^2$ we follow the prescription
from \cite{Vogt:2004dh} for the mass term and we add a contribution for the transverse momentum. Thus, 
we take:
\begin{equation}
Q^2 = (2m_c)^2 + (p_T)^2
\end{equation}
where $m_c = 1.2$ GeV/c$^2$ is the $c$ quark mass.

We study the behavior of the nuclear modification factor \RdAu\ in \dAu\ collisions
at \sqrtsNN\ = 200~GeV. \RdAu\ is presented as a function of rapidity $y$, number of collisions \Ncoll\
and transverse momentum \pT. In both shadowing cases, we consider two possibilities for the \jpsi\ transverse
momentum distributions: no transverse momentum (\pT=0) to rely on previous studies and \pT\ distribution extracted
from the measured \pp\ transverse momentum distribution in the central (\pT\ from $|y|<0.35$) and the forward 
(\pT\ from $1.2<|y|<2.2$) rapidity regions.
\begin{figure}[thb]
\includegraphics[width=\linewidth]{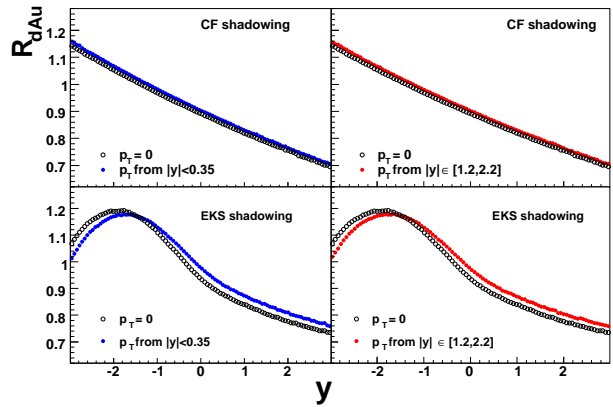}
\caption{\label{fig:results.vs.Y} 
Nuclear modification factor as a function of rapidity from CF (top) and EKS (bottom) models. Three input \pp\ \pT\ distributions
are used: \pT=0, \pT\ from $|y|<0.35$ and \pT\ from $1.2<|y|<2.2$.
}
\end{figure}

Figure~\ref{fig:results.vs.Y} shows the nuclear modification factor \RdAu\ as a function of $y$ 
for CF and EKS shadowing. In both cases the three \pT\ configurations are considered. A clear difference is 
observed between the two shadowing models at the transition between the antishadowing and the shadowing 
regions. The antishadowing region is at smaller $y$ for CF shadowing (around $-3<y<-1.5$) than for EKS shadowing 
(around $-3<y<-0.5$). As a consequence, the amount of antishadowing is larger for EKS compared to CF. Note that 
CF was originally formulated to study the amount of suppression for the nuclear ratio due to coherence in the 
shadowing region, so its kinematic relevant limits for the present energies are fixed at around $y>-2$.
The transverse momentum does not affect much the resulting \RdAu\ since the average \jpsi\ \pT\ in \pp\ 
collisions is smaller than the mass of the \jpsi\ (less than 2 GeV/c).
\begin{figure}[thb]
\includegraphics[width=\linewidth]{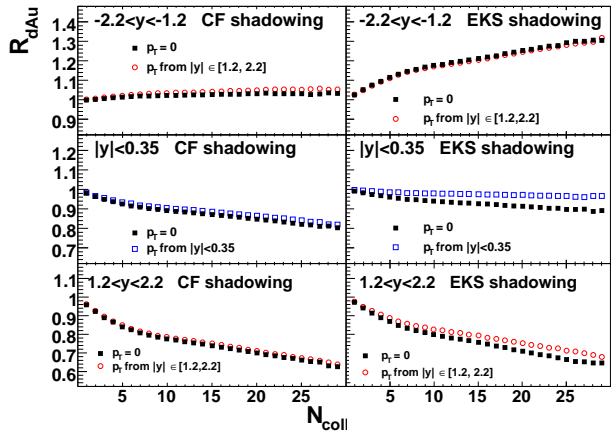}
\caption{\label{fig:results.vs.Ncoll}
Nuclear modification factor as a function of the number of collisions from CF (left) and EKS (right) models.
From up to down: backward ($-2.2<y<-1.2$), central ($|y|<0.35$) and forward ($1.2<y<2.2$) rapidity regions.
}
\end{figure}

The nuclear modification factor \RdAu\ plotted as a function of the number of collisions \Ncoll\ is shown 
in Figure~\ref{fig:results.vs.Ncoll} and leads to the same conclusions concerning the use of the real \pT\ distributions.
Even though the effects are larger for EKS shadowing, the differences between the different \pT\ configurations are pretty 
small and would probably be out of reach of experimental precision. On the other hand, the difference between the two 
shadowing models is clearly visible in the backward rapidity region ($-2.2<y<-1.2$) where EKS shadowing exhibits a 
large \RdAu\ increase while it is almost flat for the CF shadowing case. This behaviour can be directly related to 
the transition point between antishadowing and shadowing regions, as mentioned above.

\begin{figure}[thb]
\includegraphics[width=\linewidth]{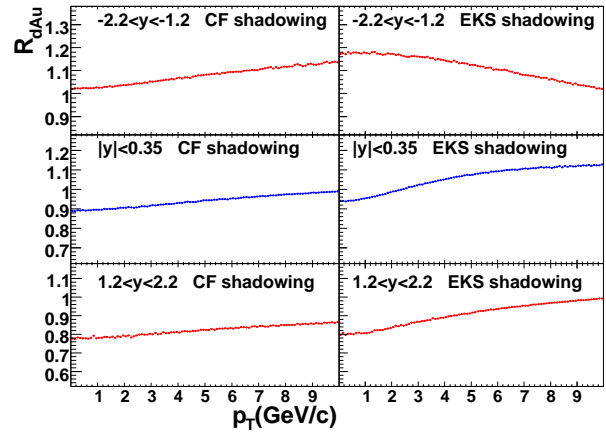}
\caption{\label{fig:results.vs.pT} 
Nuclear modification factor as a function of the \jpsi\ transverse momentum  from CF (left) and EKS (right) models.
From up to down: backward ($-2.2<y<-1.2$), central ($|y|<0.35$) and forward ($1.2<y<2.2$) rapidity regions.
}
\end{figure}

In contrast with previous figures, 
Figure~\ref{fig:results.vs.pT} clearly illustrates the gain in using non zero transverse 
momentum distributions instead of \pT=0 as done in previous studies. The \RdAu\ ratio is plotted as a function of \pT. It 
shows a dependence which can reach an amplitude of about 20~\%. The strongest \pT\ dependence is observed for the EKS 
shadowing with a specific behavior at backward rapidity ($-2.2<y<-1.2$) where \RdAu\ decreases with increasing \jpsi\ transverse 
momentum. Indeed, within EKS framework, a larger $Q^2$ in the antishadowing region leads to a smaller \RdAu\ factor. A larger \pT\ 
leads to a larger $Q^2$, so we expect a decrease of \RdAu\ in the antishadowing region and an 
increase otherwise. This effect is illustrated in Figure~\ref{fig:results.vs.Y} where the use of real \pT\ distributions induce a 
reduction of \RdAu\ for $y\leq -1.5$ while this tendency is reversed for $y\geq -1.5$. In CF shadowing, the inclusion of \pT\ in Eqs. 
(\ref{eq1}) and (\ref{eq2}) reduces the shadowing corrections, thus decreasing the function $F(y,p_T)$. It results in an increase of the 
nuclear modification factor for all rapidity regions.

In conclusion, we have compared, in the framework of a Glauber Monte Carlo, the results concerning shadowing effects from two 
different theoretical models:
CF and EKS. We have focused our attention on a new subject, the study of the shadowing dependence on $p_T$, where both models 
show a different qualitative behaviour. 
In general, by comparing the two models, we found the amount of suppression from shadowing corrections stronger in the case of 
the CF multiple scattering approach. 
However, when introducing the \pT\ contribution, a notable exception is found in the backward rapidity region where EKS shadowing show 
a stronger suppression at large \pT. 

\begin{acknowledgments}
We are thankful to A. Capella, O. Drapier and J.-P. Lansberg for fruitful discussions 
during the accomplishment of this work.
We thank J.-P. Lansberg and J. Castillo for careful 
readings of the present paper.
We acknowledge support from Ram\'on y Cajal program of Spain (E.G.F.) and IN2P3/CNRS and CEA of France.
\end{acknowledgments}

\bibliography{biblio}{}

\vskip -0.1cm
\begin{thebibliography}{8}
\expandafter\ifx\csname natexlab\endcsname\relax\def\natexlab#1{#1}\fi
\expandafter\ifx\csname bibnamefont\endcsname\relax
  \def\bibnamefont#1{#1}\fi
\expandafter\ifx\csname bibfnamefont\endcsname\relax
  \def\bibfnamefont#1{#1}\fi
\expandafter\ifx\csname citenamefont\endcsname\relax
  \def\citenamefont#1{#1}\fi
\expandafter\ifx\csname url\endcsname\relax
  \def\url#1{\texttt{#1}}\fi
\expandafter\ifx\csname urlprefix\endcsname\relax\def\urlprefix{URL }\fi
\providecommand{\bibinfo}[2]{#2}
\providecommand{\eprint}[2][]{\url{#2}}

\bibitem[{\citenamefont{Adare et~al.}(2007{\natexlab{a}})}]{Adare:2006ns}
\bibinfo{author}{\bibfnamefont{A.}~\bibnamefont{Adare}} \bibnamefont{et~al.}
  (\bibinfo{collaboration}{PHENIX}), \bibinfo{journal}{Phys. Rev. Lett.}
  \textbf{\bibinfo{volume}{98}}, \bibinfo{pages}{232301}
  (\bibinfo{year}{2007}{\natexlab{a}}), \eprint{nucl-ex/0611020}.

\bibitem[{\citenamefont{Adare et~al.}(2007{\natexlab{b}})}]{Adare:2007gn}
\bibinfo{author}{\bibfnamefont{A.}~\bibnamefont{Adare}} \bibnamefont{et~al.}
  (\bibinfo{year}{2007}{\natexlab{b}}), \eprint{arXiv:0711.3917 [nucl-ex]}.

\bibitem[{\citenamefont{Adare et~al.}(2007{\natexlab{c}})}]{Adare:2006kf}
\bibinfo{author}{\bibfnamefont{A.}~\bibnamefont{Adare}} \bibnamefont{et~al.}
  (\bibinfo{collaboration}{PHENIX}), \bibinfo{journal}{Phys. Rev. Lett.}
  \textbf{\bibinfo{volume}{98}}, \bibinfo{pages}{232002}
  (\bibinfo{year}{2007}{\natexlab{c}}), \eprint{hep-ex/0611020}.

\bibitem[{\citenamefont{Capella et~al.}(1999)\citenamefont{Capella, Kaidalov,
  and Tran Thanh~Van}}]{Capella:1999kv}
\bibinfo{author}{\bibfnamefont{A.}~\bibnamefont{Capella}},
  \bibinfo{author}{\bibfnamefont{A.}~\bibnamefont{Kaidalov}}, \bibnamefont{and}
  \bibinfo{author}{\bibfnamefont{J.}~\bibnamefont{Tran Thanh~Van}},
  \bibinfo{journal}{Heavy Ion Phys.} \textbf{\bibinfo{volume}{9}},
  \bibinfo{pages}{169} (\bibinfo{year}{1999}), \eprint{hep-ph/9903244}.

\bibitem[{\citenamefont{Capella and Ferreiro}(2005)}]{Capella:2005cn}
\bibinfo{author}{\bibfnamefont{A.}~\bibnamefont{Capella}} \bibnamefont{and}
  \bibinfo{author}{\bibfnamefont{E.~G.} \bibnamefont{Ferreiro}},
  \bibinfo{journal}{Eur. Phys. J.} \textbf{\bibinfo{volume}{C42}},
  \bibinfo{pages}{419} (\bibinfo{year}{2005}), \eprint{hep-ph/0505032}.

\bibitem[{\citenamefont{Eskola et~al.}(1999)\citenamefont{Eskola, Kolhinen, and
  Salgado}}]{Eskola:1998df}
\bibinfo{author}{\bibfnamefont{K.~J.} \bibnamefont{Eskola}},
  \bibinfo{author}{\bibfnamefont{V.~J.} \bibnamefont{Kolhinen}},
  \bibnamefont{and} \bibinfo{author}{\bibfnamefont{C.~A.}
  \bibnamefont{Salgado}}, \bibinfo{journal}{Eur. Phys. J.}
  \textbf{\bibinfo{volume}{C9}}, \bibinfo{pages}{61} (\bibinfo{year}{1999}),
  \eprint{hep-ph/9807297}.

\bibitem[{\citenamefont{Vogt}(2005)}]{Vogt:2004dh}
\bibinfo{author}{\bibfnamefont{R.}~\bibnamefont{Vogt}}, \bibinfo{journal}{Phys.
  Rev.} \textbf{\bibinfo{volume}{C71}}, \bibinfo{pages}{054902}
  (\bibinfo{year}{2005}), \eprint{hep-ph/0411378}.

\bibitem[{\citenamefont{Hodgson}(1971)}]{Hodgson:1971}
\bibinfo{author}{\bibfnamefont{P.~E.} \bibnamefont{Hodgson}},
 \bibinfo{journal}{Nuclear Reactions and Nuclear Structure, Clarendon Press},
\bibinfo{pages}{453} (\bibinfo{year}{1971}).

\bibitem[{\citenamefont{Lansberg}(2006)}]{Lansberg:2006dh}
\bibinfo{author}{\bibfnamefont{J.~P.} \bibnamefont{Lansberg}},
  \bibinfo{journal}{Int. J. Mod. Phys.} \textbf{\bibinfo{volume}{A21}},
  \bibinfo{pages}{3857} (\bibinfo{year}{2006}), \eprint{hep-ph/0602091}.

\end{thebibliography}

\end{document}